\documentclass[12pt]{article}

\usepackage{graphics,amssymb,epsfig,float}
\usepackage[usenames,dvips]{color}
\usepackage{graphicx}
\usepackage{epsfig}
\usepackage{rotating}
\usepackage{dcolumn}
\usepackage{bm}
\usepackage{cite}
\usepackage{amsmath}

\def\lsim{\;\raise0.3ex\hbox{$<$\kern-0.75em\raise-1.1ex\hbox{$\sim$}}\;}
\def\gsim{\;\raise0.3ex\hbox{$>$\kern-0.75em\raise-1.1ex\hbox{$\sim$}}\;}
\def\beq{\begin{equation}}   \def\eeq{\end{equation}}
\def\ba{\begin{array}}       \def\ea{\end{array}}
\def\bea{\begin{eqnarray}}   \def\eea{\end{eqnarray}}
\def\nn{\nonumber}

\def\k{\kappa}
\def\l{\lambda}
\def\b{\beta}

\def\noi{\noindent}

\begin{document}

\begin{titlepage}
\begin{flushright}
LPT Orsay 11-63 \\
L2C 11-\\
LUPM 11-002\\
\end{flushright}


\begin{center}
\vspace{1cm}
{\Large\bf Naturalness and Fine Tuning in the NMSSM: Implications of
Early LHC Results} \\
\vspace{2cm}

{\bf{Ulrich Ellwanger$^a$, Gr\'egory Espitalier-No\"el$^{b}$ and Cyril
Hugonie$^c$}}\\
\vspace{1cm}
\it  $^a$ LPT, UMR 8627, CNRS, Universit\'e de Paris--Sud, 91405 Orsay, France \\
\it $^b$ L2C, UMR 5221, CNRS, Universit\'e de Montpellier II, 34095 Montpellier, France \\
\it $^c$ LUPM, UMR 5299, CNRS, Universit\'e de Montpellier II, 34095 Montpellier, France \\

\end{center}
\vspace{2cm}

\begin{abstract}
We study the fine tuning in the parameter space of the
semi-constrained NMSSM, where most soft Susy breaking parameters are
universal at the GUT scale. We discuss the dependence of the fine tuning
on the soft Susy breaking parameters $M_{1/2}$ and $m_0$, and on the
Higgs masses in NMSSM specific scenarios involving large singlet-doublet
Higgs mixing or dominant Higgs-to-Higgs decays. Whereas these latter
scenarios allow {\it a priori} for considerably less fine tuning than
the constrained MSSM, the early LHC results rule out a large part of the
parameter space of the semi-constrained NMSSM corresponding to low
values of the fine tuning.
\end{abstract}

\end{titlepage}

\section{Introduction}

The first motivation for supersymmetric extensions of the Standard Model
(SM) stems from the solution of the naturalness or fine tuning problem
in the Higgs sector of the SM \cite{Witten:1981nf,Dimopoulos:1981zb,
Witten:1981kv,Kaul:1981hi,Sakai:1981gr} (besides the unification of the
gauge couplings and the possibility to explain dark matter): In the SM
with an ultraviolet cutoff $\Lambda$ much larger than the electroweak
scale $M_Z$, the bare Higgs mass squared $m_0^2$ must satisfy roughly
$m_0^2 - \Lambda^2 \sim M_Z^2$. Hence $m_0^2$ must be of the order
$\Lambda^2$, but must be finetuned relative to $\Lambda^2$ with a
precision of the order $M_Z^2/\Lambda^2$. This fine tuning is enormous
for $\Lambda$ of the order of a GUT scale, but only a tuning of ${\cal
O}(1)$ is considered as natural. Within supersymmetric (Susy) extensions
of the SM with Susy breaking terms of the order $M_\mathrm{Susy}$, the
necessary tuning between the parameters is of the order of
$M_Z^2/M_\mathrm{Susy}^2$, and hence independent from an ultraviolet
cutoff $\Lambda$.

First results of searches for Susy by the ATLAS and CMS collaborations
at the LHC, based on $\sim 1$~fb$^{-1}$ of data taken at 7~TeV
center-of-mass energy, imply lower bounds on Susy breaking gluino and
up/down squark masses in the 1~TeV range
\cite{ATL-PHYS-SLIDE-2011-352,CMS-PAS-SUS-11-003} (for the latest
publications, see the ATLAS and CMS notes on the web pages
\cite{ATLAS_twiki,CMS_twiki}). These bounds reduce the
phenomenologically viable range of parameters in Susy extensions of the
SM. However, an obviously interesting question is the impact of these
negative results on the necessary tuning between the parameters in the
remaining parameter region.

It is well known that the non-observation of a Higgs boson at LEP
\cite{Schael:2006cr} implies already a ``little fine tuning problem'' in
the Minimal Supersymmetric Standard Model (MSSM) (see e.g.
\cite{Chankowski:1997zh,Barbieri:1998uv,Kane:1998im,Giusti:1998gz}),
where the field content in the Higgs sector is as small as possible, but
large radiative corrections are required in order to lift the mass of
the lightest neutral CP-even Higgs boson above the lower LEP bound.
Large radiative corrections require relatively large Susy breaking top
squark masses compared to the electroweak scale. Via loop diagrams,
large top squark masses lead to relatively large soft Susy breaking
Higgs mass terms, which require some tuning among the parameters of the
MSSM such that the Higgs vacuum expectation values (vevs) are of ${\cal
O}(M_Z)$, well below the scale of the Higgs mass terms. The required
tuning among the parameters of the MSSM is typically of the order of a
few~\%.

The ``little fine tuning problem'' of the MSSM, originating from LEP
constraints, is alleviated in certain regions of the parameter space of
the Next-to-Minimal Supersymmetric Standard Model (NMSSM). The NMSSM is
the simplest Susy extension of the SM with a scale invariant
superpotential, i.e. where the only dimensionful parameters are the soft
Susy breaking terms. No supersymmetric Higgs mass term $\mu$ is required
as in the MSSM, since it is generated dynamically by the vacuum
expectation value of a gauge singlet superfield~$S$ (see
\cite{Maniatis:2009re, Ellwanger:2009dp} for recent reviews). Together
with the neutral components of the two SU(2) doublet Higgs fields $H_u$
and $H_d$ of the MSSM, one finds three neutral CP-even and two CP-odd
Higgs states in this model. 

The additional coupling $\l$ of $S$ to $H_u$ and $H_d$ can lead to a
larger mass $m_H$ of the SM-like neutral Higgs boson $H$, and to
mixings of the physical CP-even Higgs bosons in terms of the weak
eigenstates $S$, $H_u$ and $H_d$, implying reduced couplings of the
physical eigenstates to the $Z$ boson. Both phenomena make it easier to
satisfy the LEP bounds \cite{BasteroGil:2000bw}, and allow to alleviate
the little finetuning problem \cite{Dermisek:2007ah}. (See
\cite{Ellwanger:2006rm} for an evaluation of the upper bound on
$m_H$ of about 140~GeV if $\l$ is required to remain perturbative below
the GUT scale; if $\l$ is allowed to be larger, $m_H$ can be even larger
\cite{Barbieri:2006bg}.) 

Moreover, $H$ can decay preferably into lighter NMSSM-specific
singlet-like Higgs bosons. In this case LEP bounds on $m_H$ are lower,
and again the required fine tuning can be considerably smaller than in
the MSSM \cite{Dermisek:2005ar,Dermisek:2005gg,
Dermisek:2007yt,Dermisek:2007ah,Dermisek:2009si}. Consequently it
becomes important to study the impact of the recent bounds from the LHC
on the fine tuning within the NMSSM, which is the purpose of this paper.

The most frequently used quantitative measure $\Delta$ for fine tuning
is the maximum of the logarithmic derivative of $M_Z$ with respect to
all fundamental parameters $p_i^{\mathrm{GUT}}$ (if the fundamental
Lagrangian is given at the GUT scale) \cite{Ellis:1986yg, barbieri, de
Carlos:1993yy, Ciafaloni:1996zh, Chankowski:1997zh, Chankowski:1998xv,
Cassel:2009cx, Cassel:2010px, Cassel:2011tg,Cassel:2011zd}:
\beq\label{eq:1}
\Delta = Max\{\Delta_i^{\mathrm{GUT}}\},\qquad 
\Delta_i^{\mathrm{GUT}} = \left|\frac{\partial \ln(M_Z)}
{\partial \ln(p_i^{\mathrm{GUT}})}\right|\; .
\eeq
(See \cite{Anderson:1994dz, Chan:1997bi} for alternatives; sometimes
$M_Z^2$ instead of $M_Z$ is used in the argument of the logarithm, which
leads to an obvious additional factor of 2. Subsequently we prefer to
study linear relations between all masses and couplings.) $\Delta$
depends on the point in parameter space and is, roughly speaking,
inversely proportional to the required fine tuning (as discussed above)
between the parameters $p_i^{\mathrm{GUT}}$. Hence, for a given point in
parameter space, $\Delta$ should be as small as possible, preferably of
${\cal O}(1)$. Preferred regions in the parameter space spanned by
$p_i^{\mathrm{GUT}}$ are those where $\Delta$ is minimal (denoted by
$\Delta_{min}$). In practice, the value of $\Delta$ depends on the
choice of independent fundamental parameters defining the model, and on
the implementation of phenomenological constraints as the dark matter
relic density, the anomalous magnetic moment of the muon etc..

Including WMAP constraints on the dark matter relic density (but leaving
aside the top Yukawa coupling $h_t$ in the list of
$p_i^{\mathrm{GUT}}$), $\Delta$ has been studied recently within the
constrained MSSM (cMSSM, with universal soft Susy breaking terms at the
GUT scale) in \cite{Cassel:2009cx, Cassel:2010px, Cassel:2011tg,
Cassel:2011zd}. First investigations of  the impact of the early LHC
results on the Susy parameter space in the cMSSM have been performed in
\cite{Strumia:2011dv, Akula:2011zq,Conley:2011nn,Farina:2011bh}, in the
cMSSM with non-universal sfermion masses in \cite{Sakurai:2011pt}, and
within the general MSSM in
\cite{Scopel:2011qt,Conley:2011nn,Farina:2011bh}.

Compared to alternative procedures as likelihood scans and/or Bayesian
techniques (see \cite{LopezFogliani:2009np,Gunion:2011hs} for studies
within constrained versions of the NMSSM, and \cite{Feroz:2011bj} for a
recent discussion) a disadvantage of $\Delta$ is that it does not allow
to marginalise (i.e. to integrate over) parts of the parameters which,
in turn, would allow to determine ``most likely'' values for given
quantities as masses of specific particles, given all present
experimental constraints. (Clearly, such predictions for ``most likely''
masses seem to be of limited use; for instance, they would have failed
miserably if applied to the SM-like Higgs mass in the pre-LEP era.)

One can also leave aside the issue of quantitative fine tuning, content
oneself with the fact that the fine tuning in Susy is always much
smaller than in the SM with a large cutoff $\Lambda$, and determine
``most likely'' values for parameters exclusively from best fits to data
from electroweak precision experiments. The impact of recent LHC bounds
on such best fits has been studied recently in \cite{Allanach:2011ut,
Bechtle:2011dm,Buchmueller:2011aa, Allanach:2011wi,Bechtle:2011it,
Gunion:2011hs}.

On the other hand, the constraints from recent or future LHC results
on the quantitative fine tuning measure $\Delta$ can contribute to the
discussion on the impact of LHC results on Supersymmetry in general.
Hence we will compare these constraints within the parameter space of
the semi-constrained sNMSSM (where singlet-specific soft terms are
allowed to be non-universal) to the cMSSM, obtained from the sNMSSM in
the limit $\l,\k \to 0$. Therefore we study the dependence of $\Delta$
on the universal soft Susy breaking parameters $M_{1/2}$ and $m_0$
(gaugino and scalar masses, respectively) and on the gluino and up/down
squark masses. In particular we investigate the relevance of NMSSM
specific scenarios in the Higgs sector for fine tuning, as scenarios
with large singlet/doublet mixing and scenarios with dominant $h \to A_1
A_1$ decays \cite{Dermisek:2005ar, Dermisek:2005gg, Dermisek:2007yt,
Dermisek:2009si, Ellwanger:2009dp, Gunion:2011hs}.

In the next Section we define the model, the procedure for the
determination of $\Delta$, and discuss some specific properties of
$\Delta$ in the NMSSM. Our results and their discussion are given in
Section~3, and we conclude with a summary in Section~4.

\section{Fine Tuning in the NMSSM and the MSSM}

The NMSSM differs from the MSSM by the presence of the gauge singlet
superfield $S$. The Higgs mass term $\mu H_u H_d$ in the superpotential
$W_{MSSM}$ of the MSSM is replaced by the coupling $\lambda$ of $H_u$
and $H_d$ to $S$ and a self-coupling $\kappa S^3$, hence the
superpotential $W_{NMSSM}$ is scale invariant in this simplest
$Z_3$-invariant version of the NMSSM:
\beq\label{eq:2}
W_{NMSSM} = \lambda S H_u\cdot H_d + \frac{\kappa}{3} S^3
+h_t H_u\cdot Q\, T^c_R
+ h_b H_d\cdot Q\, B^c_R + h_\tau H_d \cdot L\, \tau^c_R\; ,
\eeq
where we have confined ourselves to the Yukawa couplings of $H_u$ and
$H_d$ to the quarks and leptons $Q$, $T_R,$ $B_R$, $L$ and $\tau_R$ of
the third generation; a sum over the three generations is implicitly
assumed. (In (\ref{eq:2}), for the first and the last time, the fields
denote superfields.) Once $S$ assumes a vev $s$, the first term in
$W_{NMSSM}$ generates an effective $\mu$-term
\beq\label{eq:3}
\mu_\mathrm{eff}=\lambda s\; .
\eeq

The soft Susy--breaking terms consist of mass terms for the
gaugino, Higgs and sfermion fields
 \bea
-{\cal L}_\mathrm{\frac12}\!&\!=\!&\! \frac{1}{2} \bigg[ 
 M_1 \tilde{B}  \tilde{B}
\!+\!M_2 \sum_{a=1}^3 \tilde{W}^a \tilde{W}_a 
\!+\!M_3 \sum_{a=1}^8 \tilde{G}^a \tilde{G}_a   
\bigg]+ \mathrm{h.c.}\; , \nn \\ 
 -{\cal L}_\mathrm{0} \!&\!=\!&\! 
m_{H_u}^2 | H_u |^2 + m_{H_d}^2 | H_d |^2 + 
m_{S}^2 | S |^2 +m_Q^2|Q^2| + m_{T}^2|T_R^2| \nn \\ &
&+\,m_B^2|B_R^2| +m_L^2|L^2|+m_\mathrm{\tau}^2|\tau_R^2|\; ,
\label{eq:4}
\eea
as well as trilinear interactions between the sfermion and the Higgs
fields, including the singlet field 
\bea
-{\cal L}_\mathrm{tril} \!&\!=\!&\! 
 \Bigl( h_t A_t\, Q\cdot H_u\, T_R^c
+ h_b  A_b\, H_d \cdot Q\, B_R^c + h_\tau A_\tau \,H_d\cdot L
\,\tau_R^c 
\nn \\ \!& &\!
+\,  \lambda A_\lambda\, H_u \cdot H_d \,S +  \frac{1}{3} \kappa 
 A_\kappa\,  S^3 \Bigl)+ \mathrm{h.c.}\;.
 \label{eq:5}
\eea
(Again, an effective MSSM-like $B$-parameter $B_\mathrm{eff}=A_\lambda
+\kappa s$ is generated.)

All parameters in the above Lagrangian depend on the energy scale via
the corresponding renormalization group (RG) equations, which account
for the dominant radiative corrections involving large logarithms. In
the constrained NMSSM, one imposes unification of the soft
Susy--breaking gaugino masses, sfermion and Higgs masses as well as
trilinear couplings at the grand unification (GUT) scale $M_{\rm GUT}$:
\bea
& & M_1  = M_2 = M_3 \equiv  M_{1/2}\, , \nn \\
& & m_{H_u}  =  m_{H_d} = m_Q = m_T = m_B = m_L
= m_{\tau} \equiv m_0\, , \nn \\
& & A_t  =  A_b = A_\tau = A_\lambda \equiv  A_0\, .
\label{eq:6}
\eea
Since the singlet superfield could play a special role (its couplings to
a hidden sector, responsible for supersymmetry breaking, could differ
from the MSSM-like fields), we will allow for non-universal
singlet-specific soft terms $m_S$ and $A_\kappa$ at the grand
unification scale. This is the so-called semi-constrained NMSSM, denoted
by sNMSSM subsequently. Including the top quark Yukawa coupling due to
its influence on the RG equations, the Lagrangian of the sNMSSM depends
on eight parameters $p_i^{\mathrm{GUT}}$ at the GUT scale:
\beq
p_i^{\mathrm{GUT}}\ = \ 
M_{1/2},  \ m_0, \ A_0, \ \lambda, \ \kappa, \
m_S, \  A_\kappa\ {\rm and}\ h_t \;.
\label{eq:7}
\eeq

The calculation of $\Delta_i^\mathrm{GUT}$ defined in (\ref{eq:1})
proceeds in two steps: First, we compute the variations
\beq\label{eq:8}
\Delta_i^{\mathrm{Susy}} = \left|\frac{\partial \ln(M_Z)}
{\partial \ln(p_i^{\mathrm{Susy}})}\right|
\eeq
with respect to the parameters $p_i^{\mathrm{Susy}}$ at the Susy scale
(the Susy scale is defined to be of the order of the soft Susy breaking
terms). Subsequently these variations are contracted with the Jacobian
\beq\label{eq:9}
J_{i j} = \left|\frac{\partial \ln(p_i^{\mathrm{Susy}})}
{\partial \ln(p_j^{\mathrm{GUT}})}\right|
\eeq
which takes care of the renormalization group running of the parameters
between the Susy and the GUT scales. Then we obtain
\beq\label{eq:10}
\Delta_j^{\mathrm{GUT}} = \sum_i  \Delta_i^{\mathrm{Susy}} J_{i j} \;.
\eeq

The parameters $p_i^{\mathrm{Susy}}$ at the Susy scale are those
appearing in the (effective) Higgs potential, whose minimization
determines the vevs $v_u$ of $H_u$, $v_d$ of $H_d$ and $s$ of $S$. These
vevs determine, in turn, the quantities
\beq\label{eq:11}
M_Z^2 = \frac{g_1^2+g_2^2}{2} (v_u^2+v_d^2), \quad \tan\beta =
\frac{v_u}{v_d}\quad {\rm and}\quad \mu_\mathrm{eff}=\lambda s\; .
\eeq

Including the dominant top quark/squark induced radiative
corrections, the three minimisation equations $E_i$ are given by
\cite{Ellwanger:2009dp}
\bea
E_1:&\ m_{H_u}^2+\mu_\mathrm{eff}^2+\l^2\,v_d^2
+\frac{g_1^2+g_2^2}{4}(v_u^2-v_d^2)
-\frac{v_d}{v_u} \mu_\mathrm{eff} (A_\l+\k s)
+\frac{3 h_t^4 v_u^2}{8\pi^2}
\ln\left(M_\mathrm{stop}^2/m_{top}^2\right) = 0\;
,\nn \\
E_2:&\ m_{H_d}^2+\mu_\mathrm{eff}^2+\l^2\,v_u^2
+\frac{g_1^2+g_2^2}{4}(v_d^2-v_u^2)
-\frac{v_u}{v_d} \mu_\mathrm{eff} (A_\l+\k s) = 0\;
,\nn \\
E_3:&\ m_{S}^2 +\k A_\k s +2\k^2 s^2 
+\l^2(v_u^2+v_d^2)-2\l\k v_u v_d -\l \frac{v_u v_d}{s} A_\l = 0\; ,
\label{eq:12}
\eea
where $M_\mathrm{stop}$ denotes an average value of the top squark
masses. (It is not necessary to be more precise here, in contrast to the
radiative corrections to the physical Higgs masses.) It is
straightforward to express the vevs $v_u$, $v_d$ and $s$ in terms of
$M_Z^2$, $\tan\beta$ and $\mu_\mathrm{eff}$ with the help of these
equations.

Hence the relevant parameters $p_i^{\mathrm{Susy}}$ at the Susy scale
are given by (leaving aside the electroweak gauge couplings $g_1$ and
$g_2$, as well as $M_\mathrm{stop}$ inside the logarithm)
\beq
p_i^{\mathrm{Susy}}\ = \ 
m_{H_u},\ m_{H_d},\ m_{S}^2,\ A_\l,\ A_\k,\ \l,\ \k,\ 
{\rm and}\ h_t \;.
\label{eq:13}
\eeq

In order to compute the variations $\Delta_i^{\mathrm{Susy}}$ (see
(\ref{eq:8})) with respect to these parameters, we use
\beq
0 = \delta E_j = \sum_i \frac{\partial E_j}{\partial
p_i^{\mathrm{Susy}}} \delta p_i^{\mathrm{Susy}} +
\frac{\partial E_j}{\partial M_Z} \delta M_Z+
\frac{\partial E_j}{\partial \tan\beta} \delta \tan\beta +
\frac{\partial E_j}{\partial \mu_\mathrm{eff}} \delta \mu_\mathrm{eff}
\label{eq:14}
\eeq
for $j = 1,2,3$. Since all partial derivatives of the equations $E_j$
can be computed explicitely, the three equations (\ref{eq:14}) can be
solved for $\delta M_Z$ (and, separately, for $\delta\tan\beta$ and
$\delta \mu_\mathrm{eff}$) as function of all $\delta
p_i^{\mathrm{Susy}}$, which allows to determine the variations
$\Delta_i^{\mathrm{Susy}}$ in (\ref{eq:8}).

At this stage it is useful to recall the origin of the ``little
fine tuning problem'' in the MSSM. Neglecting the radiative
corrections, the minimisation
equations~(\ref{eq:12}) of the Higgs potential imply,
with $\mu_\mathrm{eff} \equiv \mu$ in the MSSM,
\beq\label{eq:15}
M_Z^2 \simeq -2\mu^2 +\frac{2(m_{H_d}^2 -\tan^2\b\,
m_{H_u}^2)}{\tan^2\b -1}\; .
\eeq
In the absence of fine tuning, all terms on the right hand side of
(\ref{eq:15}) should be of comparable magnitude, and no large
cancellations should occur; hence both $\mu^2$ and $|m_{H_u}^2|$ should
not be much larger than ${\cal O}(M_Z^2)$. However, from the RG
equations one typically obtains $m_{H_u}^2 \sim -M_\mathrm{stop}^2$,
which is often required to be much larger (in absolute value) than
$M_Z^2$: At least within the MSSM, the SM-like Higgs scalar mass
increases proportionally to
$\ln\left(M_\mathrm{stop}^2/m_{top}^2\right)$ due to top/stop induced
radiative corrections. Then, large values for $M_\mathrm{stop}$ are
unavoidable in order to satisfy the LEP bound. Albeit large stop masses
are consistent with the non-observation of top squarks, they would
generate an uncomfortably large value for $-m_{H_u}^2$ which has to be
cancelled by $\mu^2$ in (\ref{eq:15}).

For large $|m_{H_u}^2| \sim \mu^2$ one finds for $\tan^2\beta \gg 1$,
following (\ref{eq:8}) with $i = m_{H_u}$ or $i =
\mu$,
\beq\label{eq:16}
\Delta_{m_{H_u}}^{\mathrm{Susy}} \sim 2\frac{|m_{H_u}^2|}{M_Z^2} \sim
\Delta_{\mu}^{\mathrm{Susy}} \sim 2\frac{\mu^2}{M_Z^2}\; .
\eeq
Accordingly large values for $\Delta_i^{\mathrm{Susy}}$ (leading,
generally, to large values for $\Delta_i^{\mathrm{GUT}}$) reflect well
the necessary fine tuning if $|m_{H_u}^2|$ and hence $\mu^2$ are large.

In the NMSSM $\mu$ is replaced by $\mu_\mathrm{eff} = \l s$.
For large $|m_{H_u}^2| \sim \mu_\mathrm{eff}^2$, the above reasoning
remains essentially unchanged: For $s \gg M_Z$ (valid in most of the
parameter space), $E_3$ in (\ref{eq:12}) gives
\beq\label{eq:17}
s \sim \frac{1}{4\kappa}  \left(-A_\kappa -
\sqrt{A_\kappa^2-8m_S^2}\right)\,.
\eeq
Replacing $\mu^2 = \mu_\mathrm{eff}^2$ and (\ref{eq:17}) for $s$ in
(\ref{eq:15}), one finds again from (\ref{eq:8})
\beq\label{eq:18}
\Delta_{m_{H_u}}^{\mathrm{Susy}} \sim 2\frac{|m_{H_u}^2|}{M_Z^2} \sim
\Delta_{\l}^{\mathrm{Susy}} \sim \Delta_{\k}^{\mathrm{Susy}} \sim 
2\frac{\mu_\mathrm{eff}^2}{M_Z^2}
\eeq in the NMSSM. Hence, quite obviously, large values for
$|m_{H_u}^2|$ are unnatural as well. However, due to the NMSSM specific
contributions to the Higgs masses and mixings or NMSSM specific Higgs
decays, LEP bounds on the Higgs sector can be satisfied for smaller
top/stop induced radiative corrections, hence for smaller values of
$M_\mathrm{stop}$, allowing for smaller values for $|m_{H_u}^2|$ and
$\mu_\mathrm{eff}^2$.

It remains to express the variations of the parameters at the Susy scale
in terms of variations of the parameters at the GUT scale, i.e. to
compute the Jacobian $J_{i j}$ in (\ref{eq:9}), via the integration of
the RG equations for the parameters. In the cMSSM (with boundary
conditions at the GUT scale as in (\ref{eq:6})) one can always write
\beq\label{eq:19}
m_{H_u}^{2} = a^{(1)}m_0^2 + a^{(2)}M_{1/2}^2
+ a^{(3)}A_0^2 + a^{(4)}M_{1/2}A_0
\eeq
and
\beq\label{eq:20}
\mu^{2} = b\ \mu_0^{2}\; ,
\eeq
where the coefficients $a^{(i)}$ and $b$ depend on the gauge and Yukawa
couplings. 

In the typical case where \emph{all} $a^{(i)}$ in (\ref{eq:19})
satisfy $|a^{(i)}| < 1$, one can verify that all variations $\partial
\ln(m_{H_u})/ \partial \ln(p_i^{\mathrm{GUT}})$, with
$p_i^{\mathrm{GUT}} = m_0,\ M_{1/2},\ A_0$, are less than 1. At first
sight, due to  $\Delta_{p_i}^{\mathrm{GUT}} <
\Delta_{m_{H_u}}$ for these parameters $p_i^{\mathrm{GUT}}$, this seems
to reduce the necessary fine tuning in the MSSM. However, due to
$\partial \ln(\mu^{2})/ \partial \ln(\mu_0^{2})\simeq 1$,
$\Delta_\mu^{\mathrm{Susy}} \simeq \Delta_\mu^{\mathrm{GUT}}$ remains
always large. Moreover, $h_t^{\mathrm{GUT}}$ should generally be
included in the list of parameters $p_i^{\mathrm{GUT}}$
\cite{Ross:1992tz,deCarlos:1993yy}, and the corresponding variation
$\Delta_{h_t}^{\mathrm{GUT}}$ can be large. This holds particularly in
the so-called focus point region of the MSSM where $m_0 \gg M_{1/2},\
A_0$, and $|a^{(1)}| \ll 1$ in (\ref{eq:19}) for specific values of
$h_t^{\mathrm{GUT}}$ (but a large derivative of $a^{(1)}m_0^2$ with
respect to $h_t^{\mathrm{GUT}}$).

In the sNMSSM, additional terms $\sim m_S^2$ and $\sim A_\k$ appear on
the right hand side of (\ref{eq:19}), which have little impact in
practice. Instead of (\ref{eq:20}), the RG equations for $\l$, $\k$ and
$m_S^2$ will now play some role since, replacing (\ref{eq:17}) for $s$
in $\mu_\mathrm{eff}$, $\mu_\mathrm{eff}$ depends on these parameters
(apart from a dependence on $A_\k$). In fact one can verify that, in the
MSSM limit of the NMSSM where $\l$, $\k \ll 1$, $\mu_\mathrm{eff}$
satisfies the \emph{same} RG equation as $\mu$. All in all we cannot
expect dramatic effects on the fine tuning from the somewhat different
running of the parameters between the Susy and the GUT scale in the
NMSSM.

In practice our computation of the different variations
$\Delta_i^\mathrm{GUT}$ in the space of parameters $p_i^\mathrm{GUT}$
(\ref{eq:7}), in order to find its maximum $\Delta$ as function of $i$
(see (\ref{eq:1})) at a specific point in the parameter space, is
performed as follows: For each such point, the code {\sf NMSPEC}
\cite{Ellwanger:2006rn} inside {\sf NMSSMTOOLs} \cite{Ellwanger:2004xm,
Ellwanger:2005dv} is used in order to compute the Higgs and sparticle
spectrum including radiative corrections as described in these
references. Constraints from LEP, B-physics and the anomalous magnetic
moment of the muon are taken care of according to the latest updates
given on the web site {\sf
http://www.th.u-psud.fr/NMHDECAY/nmssmtools.html}. (No constraints on
the dark matter relic density are imposed; however, in most cases these
could be satisfied by giving up the bino mass unification $M_1 =
M_{1/2}$ at the GUT scale, i.e. chosing an appropriate mass for the
lightest neutralino-like Susy particle (LSP) without impact on the
results relevant here.)

For phenomenologically acceptable points, $\Delta_i^\mathrm{Susy}$
is computed from the three minimization equations $E_i$ in
(\ref{eq:12}), following the procedure described above. The Jacobian
$J_{i j}$, i.e. the variations of the parameters at the Susy scale in
terms of variations of the parameters at the GUT scale, is computed 
numerically from the two loop RG equations. This allows to obtain the
necessary quantities $\Delta_i^\mathrm{GUT}$, whose maximum with respect
to all parameters $p_i^\mathrm{GUT}$ (\ref{eq:7}) defines $\Delta$.

\section{Results for the cMSSM and the sNMSSM}

\begin{figure}[hb!]
\begin{center}
\begin{tabular}{cc}
\hspace*{-5mm}
\psfig{file=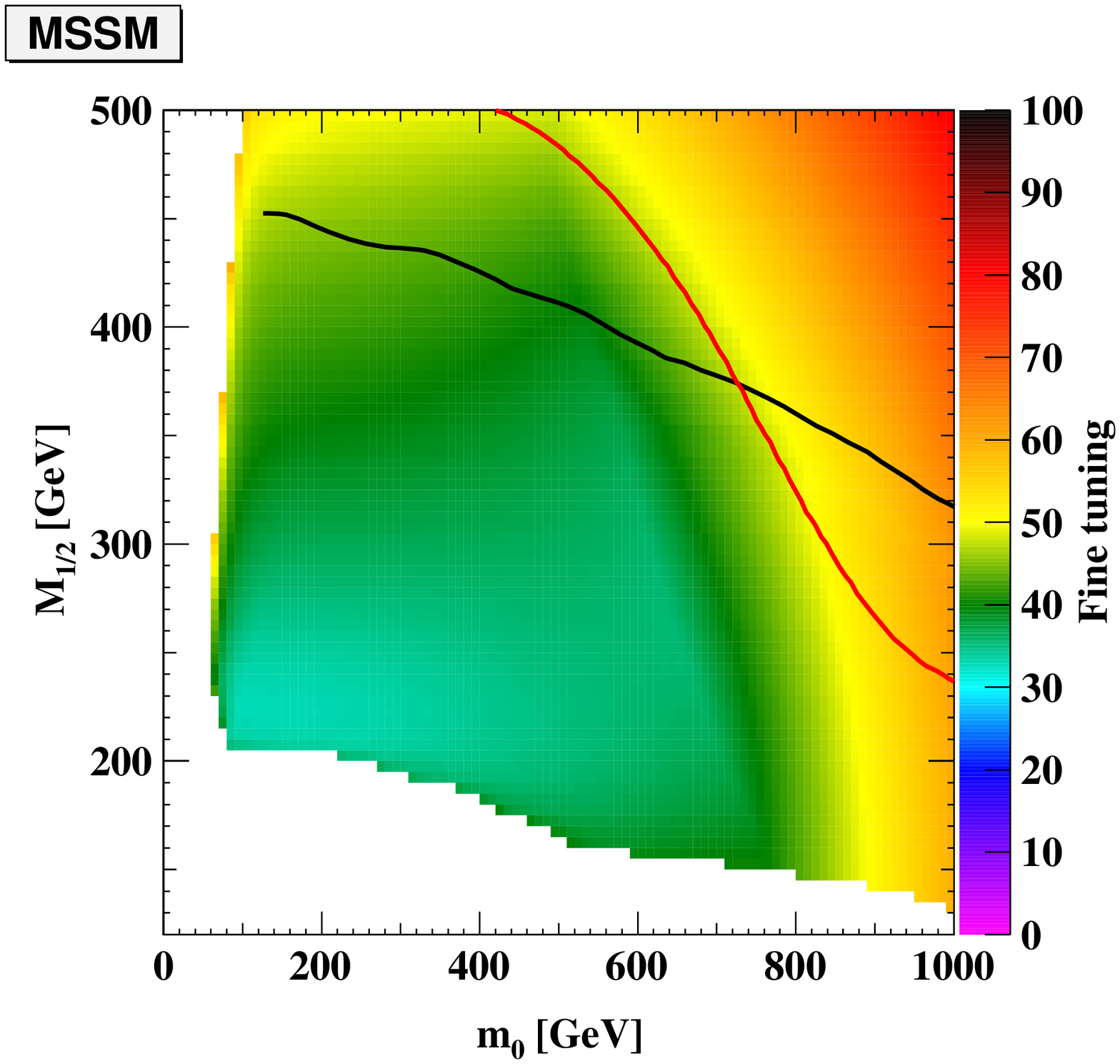, scale=0.4}
\   &
\hspace*{-5mm}
\psfig{file=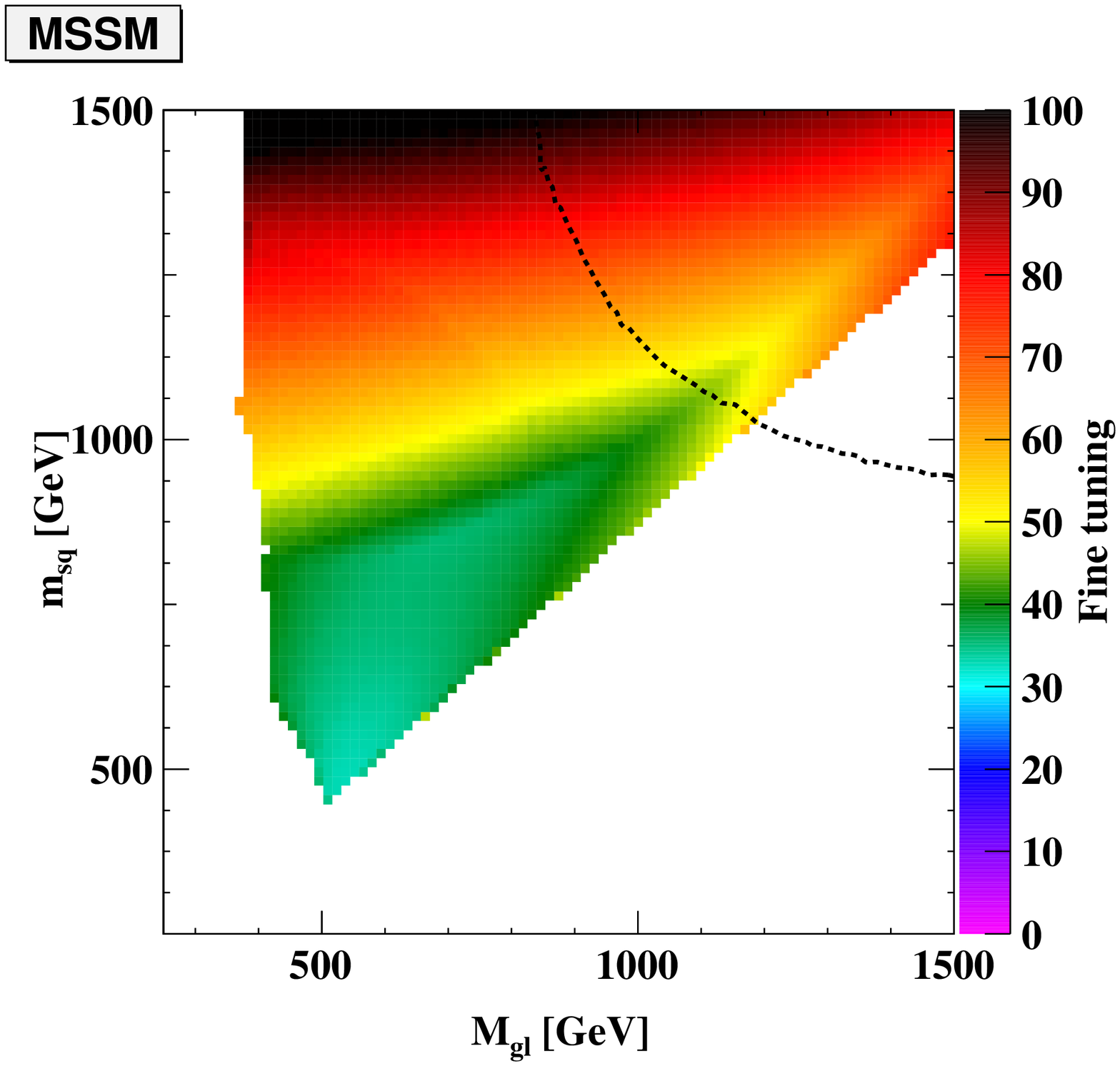, scale=0.4}
\end{tabular}
\caption{The minimal fine tuning $\Delta$ in the cMSSM. For each set of
$M_{1/2}$ and $m_0$, $\Delta$ is minimized with respect to $A_0$ and
$\tan\beta$. Left panel: $\Delta$ in the plane $M_{1/2}-m_0$. Right
panel: $\Delta$ in the plane $M_\mathrm{gluino}-m_\mathrm{squark}$.
Bounds within specific cMSSM scenarios from ATLAS
\cite{ATL-PHYS-SLIDE-2011-352} are indicated as black lines, and from
CMS \cite{CMS-PAS-SUS-11-003} as red lines (see text).}
\label{fig:1}
\end{center}
\end{figure}

To start with, we apply our procedure to the cMSSM, which allows for
comparisons with the sNMSSM and the available literature. The relevant
parameters $p_i^{\mathrm{GUT}}(\mathrm{cMSSM})$ at the GUT scale are
\beq
p_i^{\mathrm{GUT}}(\mathrm{cMSSM})\ = \ 
M_{1/2},  \ m_0, \ A_0, \ \mu_0, \ B_0, \ {\rm and}\ h_t \;.
\label{eq:21}
\eeq
As usual, $\mu$ and $B$ (and hence $\mu_0$ and $B_0$) are determined by
$M_Z$ and $\tan\beta$, but they still contribute to the definition of
$\Delta$. Next we scan over a grid of values of $M_{1/2}$ and $m_0$. For
each set of these values, we scan over $A_0$ and $\tan\beta$. Keeping
only sets of parameters consistent with all phenomenological
constraints, we look for values of $A_0$ and $\tan\beta$ which minimize
$\Delta$ as defined in (\ref{eq:1}) for fixed $M_{1/2}$ and $m_0$. The
resulting minimal values of $\Delta$ can be represented in the plane
$M_{1/2}-m_0$, or in the plane $M_\mathrm{gluino}-m_\mathrm{squark}$
(where $m_\mathrm{squark}$ refers to squarks of the first generation) in
Figs.~\ref{fig:1}. 

In order to guide the eye, we have indicated lower bounds from ATLAS and
CMS notes on analyses of jets and missing $E_T$, based on an integrated
luminosity $L^{int} \simeq 1$~fb$^{-1}$: In the plane $M_{1/2}-m_0$,
lower bounds from ATLAS \cite{ATL-PHYS-SLIDE-2011-352} (interpreted
within the cMSSM with $\tan\beta = 10$, $A_0 = 0$) are shown as a black
line, and lower bounds from CMS \cite{CMS-PAS-SUS-11-003} are shown as a
red line. In the plane $M_\mathrm{gluino}-m_\mathrm{squark}$, lower
bounds from ATLAS \cite{ATL-PHYS-SLIDE-2011-352} are shown as as a black
line. (The latter bounds from ATLAS are obtained in a simplified model
where squarks decay only into quarks + a neutralino with a branching
ratio of 100~\%.)

In the white regions in Figs.~\ref{fig:1}, phenomenological constraints
cannot be satisfied for any values of $A_0$ and $\tan\beta$: Either a
stau would be the LSP (left hand side of the left panel), or a charged
slepton, chargino, neutralino or a CP-even Higgs boson is excluded by
LEP2/Tevatron (lower part of the left panel), or squarks are excluded by
CDF/D0 (left hand side of the right panel) or are theoretically
unaccessible (right hand side of the right panel).

\begin{figure}[ht!]
\begin{center}
\vspace*{10mm}
\psfig{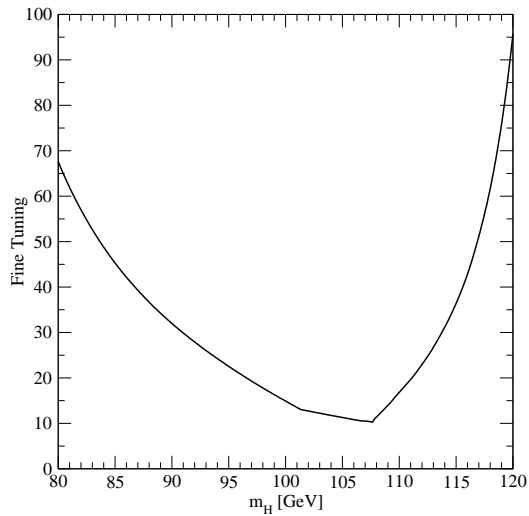}
\vspace*{5mm}
\caption{$\Delta$ as function of the SM-like Higgs mass $m_H$ in the
cMSSM, without imposing LEP constraints on the Higgs sector.}
\label{fig:2}
\end{center}
\end{figure}

Note that, for $M_{1/2} \lsim 350$~GeV and $m_0 \lsim 700$~GeV in the
left panel, $\Delta$ decreases hardly with decreasing Susy breaking
parameters $M_{1/2}$ and $m_0$; the minimal value of $\Delta$ in the
pre-LHC allowed region is about $\sim 33$. In this region of the
parameter space we observe the ''little fine tuning problem'' of the
MSSM due to the LEP bound on the SM-like Higgs mass, which hardly
depends on $M_{1/2}$ and $m_0$. The impact of the LEP bound becomes
clear once we minimize the fine tuning $\Delta$ for fixed lightest Higgs
mass $m_H$ (without imposing LEP constraints on the Higgs sector) as a
function not only of $A_0$ and $\tan\beta$, but also of $M_{1/2}$ and
$m_0$. The result is shown in Fig.~\ref{fig:2}.

We see the strong increase of $\Delta$ with $m_H$ for $m_H \gsim
108$~GeV. Accordingly the LEP constraint $m_H \gsim 114$~GeV implies
$\Delta \gsim 33$ in agreement with Figs.~\ref{fig:1} (for low values of
$M_{1/2}$ and $m_0$). In fact, $m_H$ is always just above 114~GeV in the
entire planes in Fig.~\ref{fig:1}, once $A_0$ and $\tan\beta$ are chosen
such that $\Delta$ is minimized for fixed $M_{1/2}$ and $m_0$, but LEP
constraints are applied. 

Hence, for low values of $M_{1/2}$ and $m_0$ (in the region $M_{1/2}
\lsim 350$~GeV and $m_0 \lsim 700$~GeV), $\Delta$ is determined by the
LEP constraints on $m_H$ (implying the little fine tuning problem of the
MSSM) and not much affected by the lower bounds on gluino and squark
masses from early LHC searches: In the region above the ATLAS/CMS
bounds, the minimal value of $\Delta$ increases only to $\sim 40-50$. 

We also see in Fig.~\ref{fig:2} that $\Delta$ does not decrease
systematically with $m_H$, once lower bounds on sparticle masses from
LEP and the Tevatron are imposed: In order to minimize $\Delta$,
preferred values of $m_H$ would be in the range $100 - 110$~GeV which is
excluded in the MSSM, but not in the NMSSM (see below).

For larger values of $M_{1/2}$ or $m_0$, the origin of the required
fine tuning is different: Here it is simply the fact that the weak scale
(determined essentially by $-2(\mu^2+m_{H_u}^2)$) is
small compared to the Susy breaking scale, which requires some tuning
between the parameters. Since $m_{H_u}$ at the Susy scale is closely
related to the squark masses, $\Delta$ increases rapidly with
$m_{squark}$ (for $m_{squark}\gsim 1$~TeV) as it is visible in the right
panel in Figs.~\ref{fig:1}.

The fine tuning in the cMSSM has recently been analysed in
\cite{Cassel:2009cx, Cassel:2010px, Cassel:2011tg, Cassel:2011zd}. The
procedure and precision in these papers is similar to ours, except that
constraints on the dark matter relic density are applied in
\cite{Cassel:2009cx, Cassel:2010px, Cassel:2011tg, Cassel:2011zd}, but
contributions to $\Delta$ from the top Yukawa coupling $h_t$ are left
aside. From Fig.~7d in \cite{Cassel:2010px} we find, once LEP
constraints are applied, a minimal value of $\Delta \sim 70$ for not too
large values of $m_0$. Given that $\Delta$ in \cite{Cassel:2010px} is
twice as large as our $\Delta$ defined in (\ref{eq:1}), this coincides
well with the left panel of Figs.~\ref{fig:1} for moderate values of
$M_{1/2}$. However, for $m_0 \gsim 800$~GeV, $\Delta$ decreases to $\sim
10$ in \cite{Cassel:2010px}, whereas $\Delta$ increases with $m_0$ in
the left panel of Figs.~\ref{fig:1}. In fact, for larger values of
$m_0$, $\Delta$ is dominated by contributions from $h_t$ whose absence
in \cite{Cassel:2009cx, Cassel:2010px, Cassel:2011tg, Cassel:2011zd}
explains the different results for $\Delta$ in this region.

Next we turn to the sNMSSM. In various regions of the parameter space of
the sNMSSM, unconventional properties of the Higgs sector allow to
alleviate the LEP bounds, lowering the minimal possible values of
$\Delta$. We found it interesting to study these lower bounds on
$\Delta$ separately for different scenarios in the NMSSM Higgs sector
(see also \cite{Dermisek:2007ah}), since these will have very different
implications for future Higgs searches at the LHC. Hence we distinguish
subsequently the following two scenarios:

\vspace{5mm}
(1) The lightest CP even Higgs boson $H_1$ has a large singlet component
(H/S mixing). This implies a reduced coupling to the $Z$ boson, and
allows for $H_1$ masses well below 114~GeV \cite{Schael:2006cr}.

\vspace{5mm}
(2) A CP-even Higgs boson $H$ decays dominantly into a pair of lighter
CP-odd bosons, $H \to AA$, allowing again for $H_1$ masses well below
114~GeV \cite{Schael:2006cr}. (We omit the index $1$ of $A_1$ in the
following.)
\vspace{5mm}

The search for the minimal fine tuning $\Delta$ in each of these
scenarios is performed similar to the procedure in the cMSSM: Again we
scan over a grid of values of $M_{1/2}$ and $m_0$. Now, for each set of
these values, we scan over $A_0$, $\tan\beta$, $\lambda$ and $A_\kappa$
($\kappa$ and $m_S$ are determined by $M_Z$ and $\tan\beta$, but
included in the definition of $\Delta$) using a Monte Carlo Markov Chain
(MCMC) technique. Keeping only sets of parameters consistent with all
phenomenological constraints, we look for values of $A_0$, $\tan\beta$,
$\lambda$ and $A_\kappa$ which minimize $\Delta$ as defined in
(\ref{eq:1}) for fixed $M_{1/2}$ and $m_0$, allowing us to represent the
resulting minimal values of $\Delta$ in the plane $M_{1/2}-m_0$, or in
the plane $M_\mathrm{gluino}-m_\mathrm{squark}$. (In order to
distinguish the scenarios above, we require essentially $BR(H_1 \to bb)
> 0.7$ for scenario (1), but $BR(H_1 \to AA) > 0.2$ for scenario (2).)

\begin{figure}[ht!]
\begin{center}
\begin{tabular}{cc}
\hspace*{-5mm}
\psfig{file=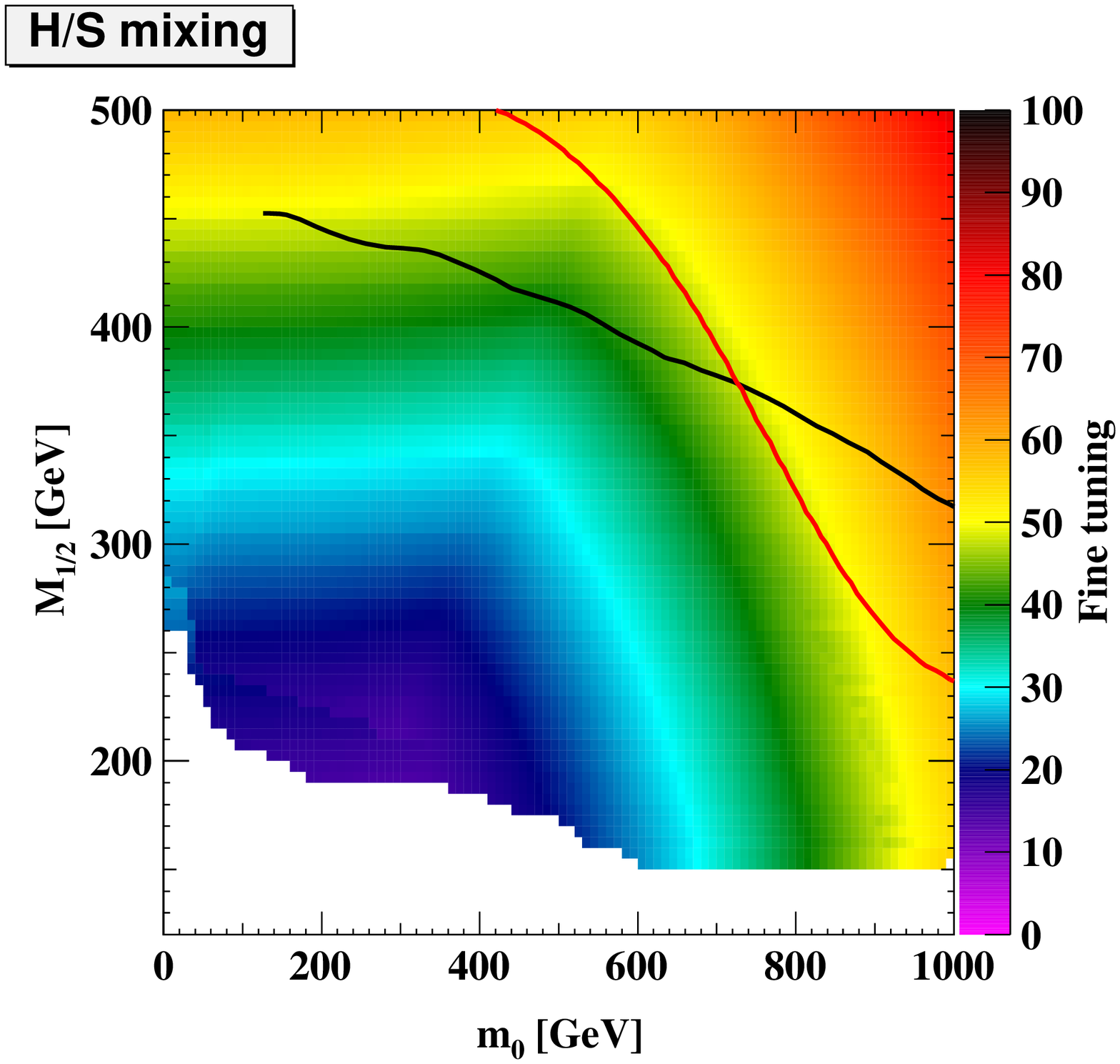, scale=0.4}
\  &
\hspace*{-5mm}
\psfig{file=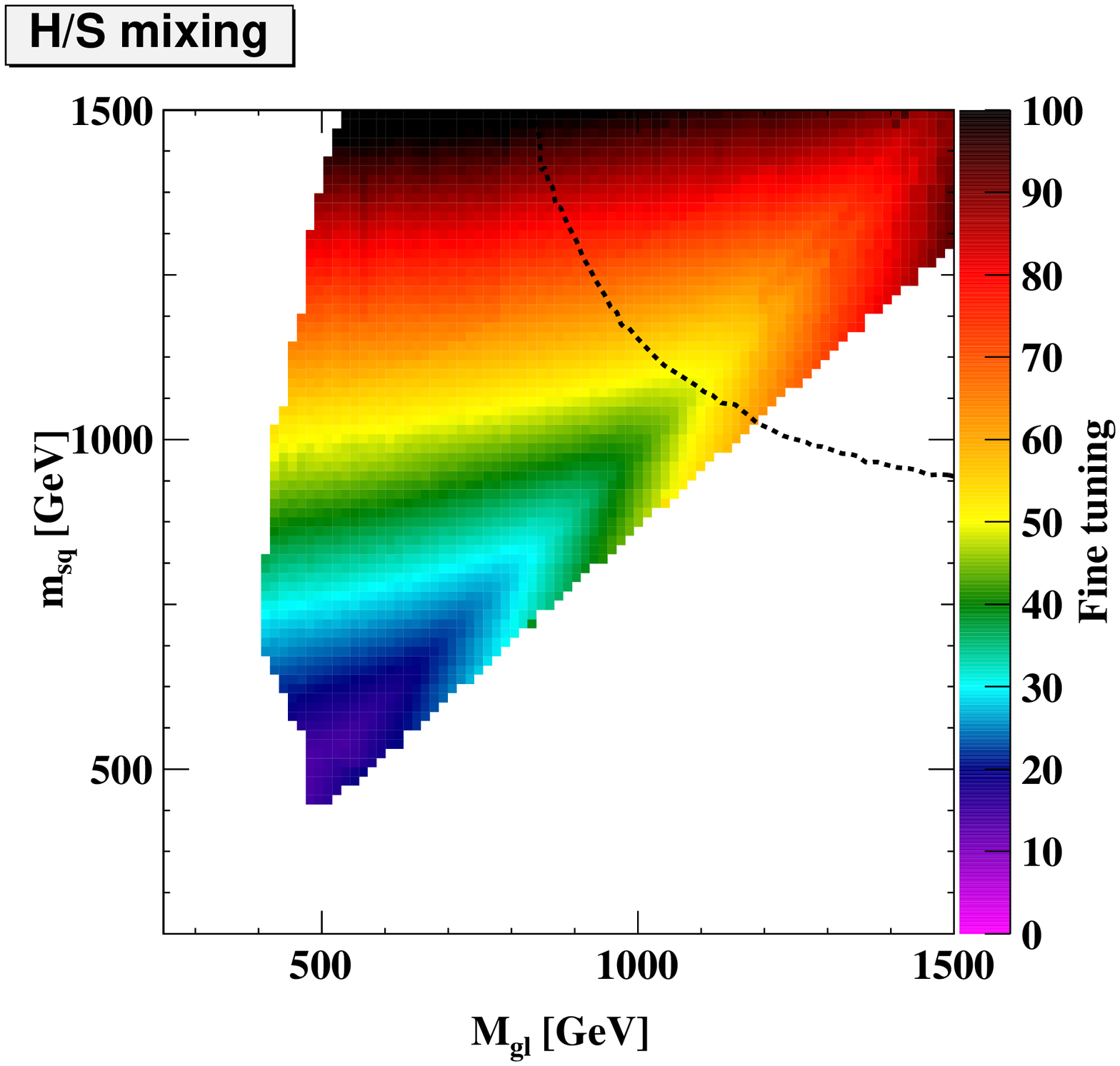, scale=0.4}
\end{tabular}
\caption{The minimal fine tuning $\Delta$, defined as in
Figs.~\ref{fig:1} for the cMSSM, in the sNMSSM scenario (1).}
\label{fig:3}
\end{center}
\end{figure}

In the scenario (1) (H/S mixing), the corresponding results for $\Delta$
are shown in Figs.~\ref{fig:3}. Now the constraint on the left hand side
in the left panel from the absence of a stau LSP has disappeared, since
a singlino-like neutralino can be the LSP. From here onwards, the bounds
from ATLAS and CMS are indicative only, since the signals for
supersymmetry in the NMSSM can be different notably in the case
of a singlino-like LSP \cite{Ellwanger:2010es}.

Compared to the fine tuning in the cMSSM in Figs.~\ref{fig:1}, we see
that $\Delta$ can be considerably smaller for not too large values of
the Susy breaking parameters $M_{1/2}$ and $m_0$ \cite{Dermisek:2007ah}:
A large singlet
component of $H_1$ (in the range $0.8 - 0.85$) allows for lighter $H_1$
masses compatible with LEP constraints, which reduces the required fine
tuning, see Fig.~\ref{fig:4} below. The mass of the second mostly
SM-like CP-even Higgs boson $H_2$ is always just above 114~GeV. This is
now easier to satisfy than in the MSSM, since the doublet/singlet mixing
shifts the mass of the mostly doublet-like Higgs boson upwards. The
values of the NMSSM-specific coupling $\l$ do not have to be large to
this end; its value is always $\lsim 0.01$.

Respecting just the pre-LHC phenomenological constraints,
the fine tuning measure $\Delta$ can be as small as 14 for low values of
$M_{1/2}$ and $m_0$ in this scenario, but this is precisely the region
which is constrained by the first unsuccessful searches for Susy at the
LHC. (As stated above, these constraints depend on the decay properties
of the u/d-squarks and gluinos. Additional bino $\to$ neutralino decay
processes can reduce $E_T^{miss}$ signature in all sparticle decay
cascades. Applying nevertheless the bounds for the cMSSM scenarios
studied by the ATLAS and CMS collaborations to the sNMSSM, we find that
the smallest admissible value of $\Delta$ becomes $\sim 44$ for $M_{1/2}
\sim 400$~GeV, $m_0 \sim 600$~GeV, similar to the smallest admissible
value of $\Delta$ in the cMSSM.)

For larger values of $M_{1/2}$ and $m_0$, the fine tuning within this
scenario (1) of the sNMSSM becomes similar to the one within the cMSSM:
As explained above, the origin of the fine tuning is now the smallness
of the weak scale with respect to $M_{Susy}$ and not the LEP constraints
on the Higgs mass; hence the possibility to alleviate the LEP
constraints within the sNMSSM is less relevant.

In Fig.~\ref{fig:4} we show $\Delta$ as function of the mass $m_{H_1}$
of the dominantly singlet-like Higgs state, minimizing $\Delta$
as a function of $m_0$, $M_{1/2}$, $A_0$, $\tan\beta$ taking into
account LEP constraints from Higgs and sparticle searches.
(The irregular structures originate from the LEP constraints on $H_1 \to
b\bar{b}$, which lead to irregular upper bounds on the coupling of $H_1$
to the $Z$ boson as function of $m_{H_1}$.) Again, $\Delta$ does not
decrease systematically with decreasing $m_{H_1}$ given the
phenomenological constraints on sparticle masses, but the behaviour is
somewhat different from the dependence of $\Delta$ on the SM-like Higgs
mass in the MSSM. Here we find that $\Delta$ is minimal for $m_{H_1}$
below $\sim 100$~GeV which coincides with the Higgs mass range where a
2.3~$\sigma$ excess in $H \to b\bar{b}$ is observed at LEP~2
\cite{Schael:2006cr}. (This excess could be explained here since,
although $H_1$ is dominantly a gauge singlet, it still has a
nonvanishing -- but reduced -- coupling to the $Z$ boson, and will decay
dominantly into $b\bar{b}$.)

\begin{figure}[ht!]
\begin{center}
\vspace*{1cm}
\psfig{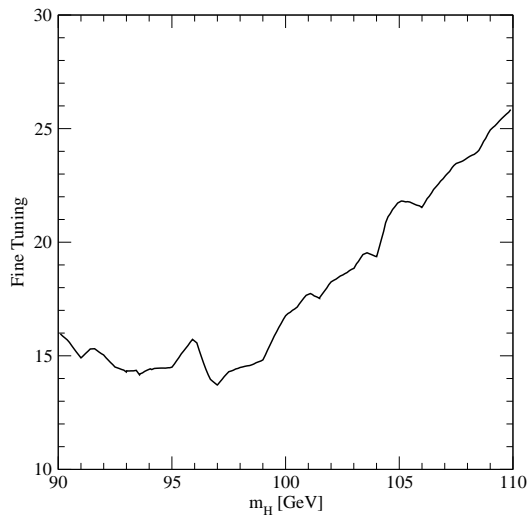}
\vspace*{5mm}
\caption{$\Delta$ as function of the mass $m_{H_1}$ of the dominantly
singlet-like Higgs state in the mixing scenario~(1) described above,
including constraints from LEP.}
\label{fig:4}
\end{center}
\end{figure}

\newpage
Next we turn to the sNMSSM scenario (2) involving light pseudoscalars
$A$, into which the SM-like CP-even Higgs boson $H$ can decay. Again the
required fine tuning is reduced \cite{Dermisek:2005ar, Dermisek:2005gg},
since LEP constraints allow for smaller $H$ masses than 114~GeV. These
constraints depend on $m_A$:

\noi a) For $m_A \gsim 10.5$~GeV, $A$ will dominantly decay into a pair
of $b\bar{b}$ quarks. This signature has been studied by the OPAL and
DELPHI groups at LEP \cite{Abbiendi:2004ww,Abdallah:2004wy} implying
$m_H \gsim 105-110$~GeV if $H$ has SM-like couplings to the $Z$ boson
\cite{Schael:2006cr}. Still, this lower bound on $m_H$ allows for lower
values of $\Delta$.

\noi b) For $m_A \lsim 10.5$~GeV, $A$ decays dominantly into a pair of
$\tau$ leptons. The signature $H \to AA \to 4\,\tau$ has recently been
re-analysed by the ALEPH group \cite{Schael:2010aw} for $m_A < 12$~GeV,
implying again lower limits on $m_H$. However, for $m_A$ in the range
$9\ \mathrm{GeV} \lsim m_{A} \lsim 10.5$~GeV, $A$ can and will mix
strongly with the CP-odd $\eta_b(nS)$ states \cite{Domingo:2009tb} which
implies a considerable reduction of the $BR(A \to \tau^+ \tau^-)$
\cite{Domingo:2011rn}. Hence bounds on $H \to AA \to 4\,\tau$ hardly
constrain $m_H$ in this case; now we find that the dominant constraints
result from the lower limits on $m_H$ depending on remaining
sub-dominant $BR(H \to b\bar{b})$ \cite{Schael:2006cr}.

\begin{figure}[ht!]
\begin{center}
\begin{tabular}{cc}
\hspace*{-5mm}
\psfig{file=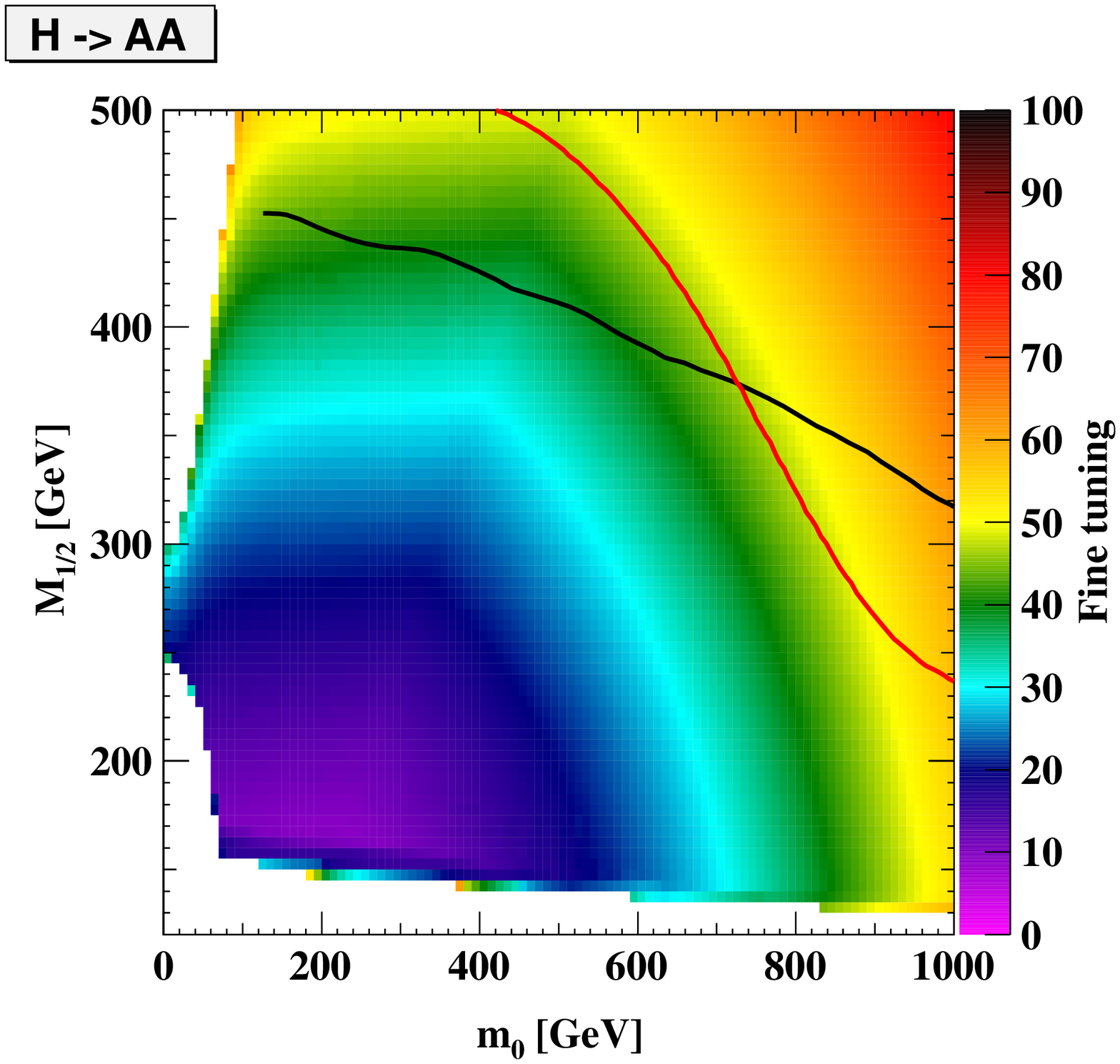, scale=0.4}
\ &
\hspace*{-5mm}
\psfig{file=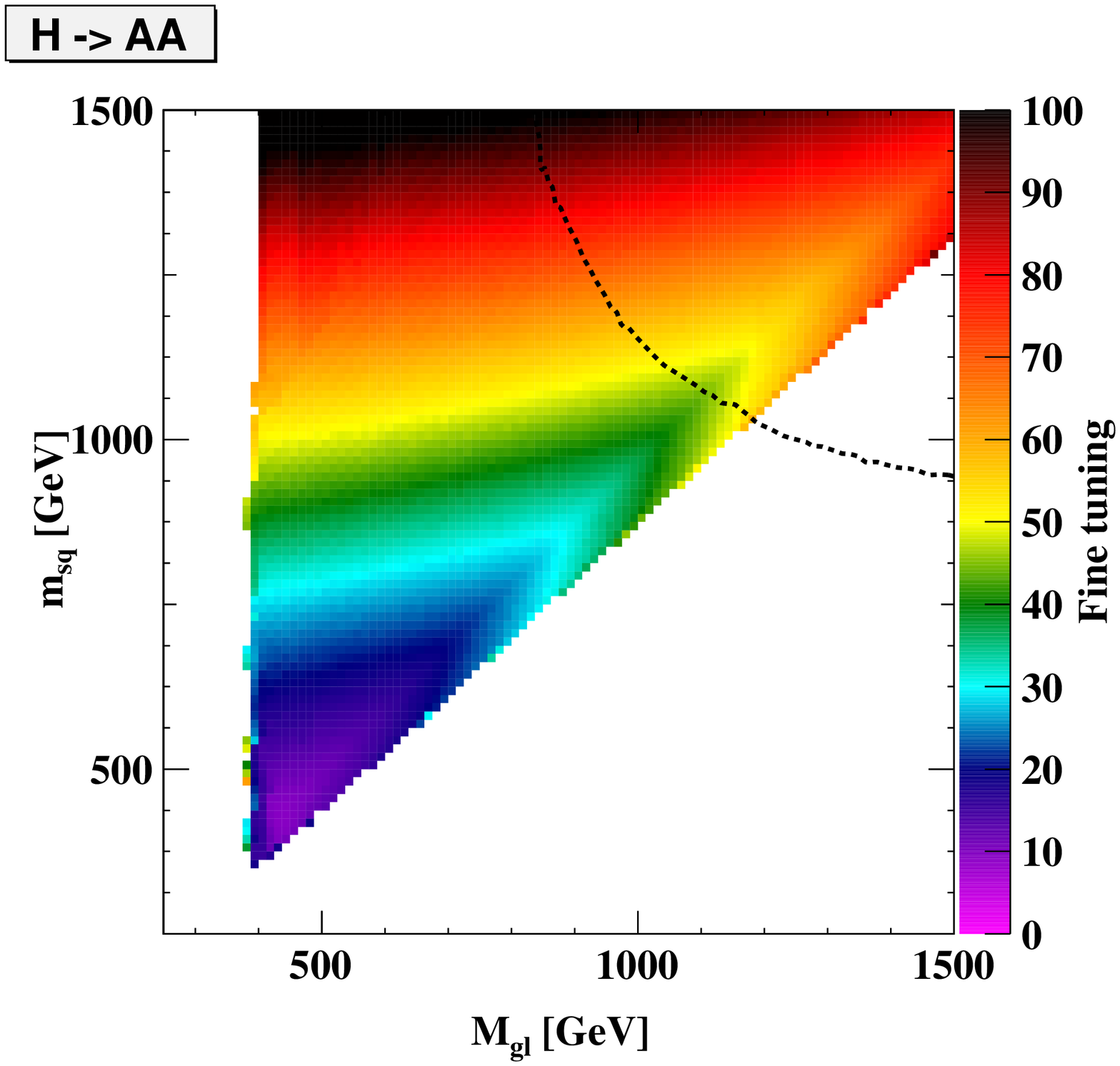, scale=0.4}
\end{tabular}
\caption{The minimal fine tuning $\Delta$, defined as in
Figs.~\ref{fig:1} for the cMSSM, in the sNMSSM scenario~(2).}
\label{fig:5}
\end{center}
\end{figure}

Our results for the minimal fine tuning $\Delta$ in the sNMSSM scenario
(2) are shown in Figs.~\ref{fig:5} in the same planes as before. Similar
to the sNMSSM scenario (1), respecting just the pre-LHC constraints, the
fine tuning measure $\Delta$ can be as small as 9 for low values of
$M_{1/2}$ and $m_0$. Applying naively the bounds for the cMSSM
scenarios studied by the ATLAS and CMS collaborations to the sNMSSM, we
find that the smallest admissible value of $\Delta$ becomes $\sim 39$
for $M_{1/2} \sim 375$~GeV, $m_0 \sim 700$~GeV, hence below the smalles
admissible value in the cMSSM.

In the region of low $\Delta$, the $BR(H \to AA)$ is larger than 80~\%.
Again, the values of the NMSSM-specific coupling $\l$ do not have to be
large in order to favour $H \to AA$ decays once these are kinematically
allowed; the value of $\l$ varies between 0.02 and 0.16 in
Figs.~\ref{fig:5}.

\begin{figure}[ht!]
\begin{center}
\vspace*{10mm}
\psfig{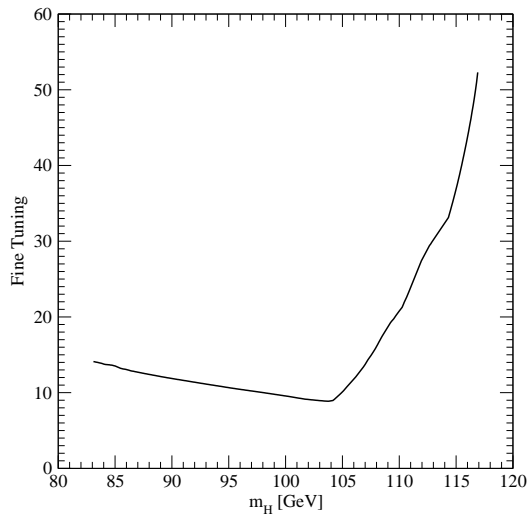}
\vspace*{5mm}
\caption{$\Delta$ as function of $m_{H}$ in the scenario~(2), once LEP
constraints are imposed.}
\label{fig:6}
\end{center}
\end{figure}

In Fig.~\ref{fig:6} we show the minimal value of $\Delta$ as function of
the mass $m_{H}$ of the now SM-like state~$H$, imposing LEP constraints.
Again the preferred value of $m_{H}$ is not always as small as possible;
now the minimal fine tuning is obtained for $m_{H}$ in the range
$100-105$~GeV, and increases again strongly for larger values of $m_H$.

\begin{figure}[ht!]
\begin{center}
\vspace*{10mm}
\psfig{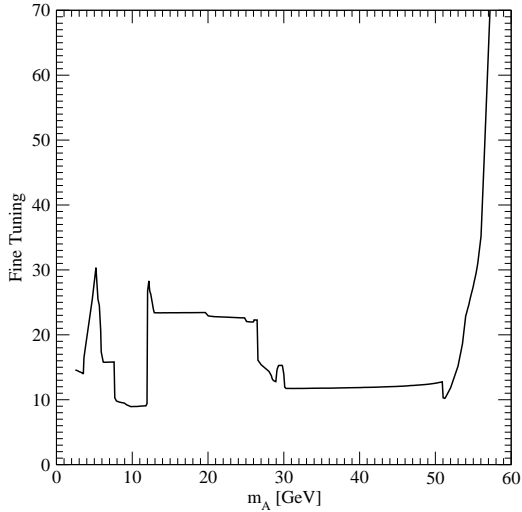}
\vspace*{5mm}
\caption{$\Delta$ as function of $m_{A}$ in the scenario~(2), once LEP
constraints are imposed.}
\label{fig:7}
\end{center}
\end{figure}

It is also interesting to study the dependence of the fine tuning on
$m_A$. At first sight, $\Delta$ depends hardly on $m_A$ within the
sNMSSM, where $A_\k$ differs from $A_0$ at the GUT scale: A variation of
$m_A$ from $0$ to $\sim 50$~GeV (still allowing for $H \to AA$ decays)
corresponds to a variation of $|A_\k|$ from $\sim 5$~GeV to $\sim
40$~GeV at the GUT scale, which has practically no effect on the fine
tuning measure $\Delta$. However, different values for $m_A$ correspond
to different lower LEP bounds on $m_H$; lower bounds $m_H$, in turn,
affect $\Delta$ as in Fig.~\ref{fig:6}. The resulting minimal values of
$\Delta$ as function of $m_A$ are shown in Fig.~\ref{fig:7}.

The structures in Fig.~\ref{fig:7} can be explained as follows: For
$m_A \gsim 11$~GeV, LEP bounds on $H_2 \to H_1 H_1 \to 4b$ (here with $H_2
\sim H$, and $H_1 \sim A$) apply, see Figs.~3 in \cite{Schael:2006cr}. The
corresponding lower bounds on $m_H$ are somewhat weaker for $m_A \gsim
30$~GeV than for $11\ \mathrm{GeV}\lsim m_A \lsim 30$~GeV, leading to
a lower fine tuning for $m_A \gsim 30$~GeV where $m_H \gsim 106$~GeV.
(For $m_A \sim 51$~GeV, the lower bound on $m_H$ has a dip implying a
dip in $\Delta$. For $m_A \sim 57$~GeV, $\Delta$ increases since
$m_H > 2 m_A > 114$~GeV is required for kinematic reasons; here this
scenario is obviously not preferred.)
For $m_A \lsim 11$~GeV, strong lower bounds on $m_H$ from ALEPH
\cite{Schael:2010aw} on the $BR(H \to A A \to 4\tau)$ apply at first
sight. However, due to a reduced branching ratio for $A \to 2\tau$ form
$A-\eta_b$ mixing \cite{Domingo:2011rn}, the window $9\
\mathrm{GeV}\lsim m_A \lsim 11$~GeV is hardly affected by these
constraints, allowing for $m_H$ to assume values for which $\Delta$ is
minimal according to Fig.~\ref{fig:6}. (For $m_A \sim 5$~GeV, parameters
have to be tuned in order to satisfy constraints from $B_s \to
\mu^+\mu^-$.) Hence we find two distinct
regions for $m_A$ where $H \to AA$ decays allow for a considerable
reduction of the fine tuning.

\section {Conclusions}

We have studied the amount of fine tuning in the parameter
space of the semi-constrained NMSSM, and compared it to the cMSSM.
Representing the minimal fine tuning in the plane $M_{1/2} - m_0$ allowed
us to study the impact of early LHC results. First we verified
quantitatively, to what extend the NMSSM-specific scenarios in the Higgs
sector allow to alleviate the LEP constraints with respect to the MSSM.
We found indeed that a considerable reduction of the fine tuning is
possible in scenarios where the lightest CP-even Higgs state is
dominantly singlet like, or in scenarios where $H \to AA$ decays are
possible, notably for $m_A \sim (10 \pm 1)$~GeV and $30\ \mathrm{GeV}
\lsim m_A \lsim 55\ \mathrm{GeV}$.

If one applies naively the early LHC constraints on the cMSSM in the
$M_{1/2}-m_0$-plane to the sNMSSM, the impact on the minimal fine tuning
is stronger in the sNMSSM since these constraints affect in particular
the region of smaller Susy breaking terms where the fine tuning in the
sNMSSM can be relatively small. Still, the necessary fine tuning in the
sNMSSM in scenarios where $H \to AA$ decays are possible is smaller than
in the cMSSM, but only if the Susy breaking terms
are not too large (hence if sparticles are not too heavy). Otherwise, if
the Susy breaking terms are larger, the fine tuning does not originate
from LEP constraints on the Higgs sector, but from the smallness of the
weak scale with respect to the Susy scale. This hardly depends on
details of the Higgs sector of the Susy model, as long as the
fundamental parameters are defined at the GUT scale and influence each
other through the renormalization group running between the GUT and the
weak scale. Hence the same question should be re-analysed in the NMSSM
with gauge mediated supersymmetry breaking \cite{Ellwanger:2008py},
where the Susy breaking parameters can originate from much lower
scales, and where the Susy breaking Higgs mass terms can differ
considerably from squark mass terms.

\section*{Acknowledgements}

U.~E. acknowledges support from the French ANR LFV-CPV-LHC.
C.~H. acknowledges support from the French ANR TAPDMS 
(ANR-09-JCJC-0146).



\begin{thebibliography}{99}
  
\bibitem{Witten:1981nf}
  E.~Witten,  Nucl.\ Phys.\  B {\bf 188} (1981) 513.

\bibitem{Dimopoulos:1981zb}
  S.~Dimopoulos and H.~Georgi,
  Nucl.\ Phys.\  B {\bf 193} (1981) 150.  
   
\bibitem{Witten:1981kv}
  E.~Witten,
  Phys.\ Lett.\  B {\bf 105} (1981) 267.
   
\bibitem{Sakai:1981gr}
  N.~Sakai,  Z.\ Phys.\  C {\bf 11} (1981) 153.

\bibitem{Kaul:1981hi}
  R.~K.~Kaul and P.~Majumdar,
  Nucl.\ Phys.\  B {\bf 199} (1982) 36.

\bibitem{ATL-PHYS-SLIDE-2011-352} ATLAS Collaboration,
ATL-COM-PHYS-2011-981, presentation at the
2011 Europhysics Conference On High Energy Physics, Grenoble,
France, 21 - 27 July 2011

\bibitem{CMS-PAS-SUS-11-003} CMS Collaboration,
``Search for supersymmetry in all-hadronic events with $\alpha_T$'',
CMS-PAS-SUS-11-003, presentation at the
2011 Europhysics Conference On High Energy Physics, Grenoble,
France, 21 - 27 July 2011

\bibitem{ATLAS_twiki}
{\sf 
https://twiki.cern.ch/twiki/bin/view/AtlasPublic/SupersymmetryPublicResults}

\bibitem{CMS_twiki}
{\sf
https://twiki.cern.ch/twiki/bin/view/CMSPublic/PhysicsResultsSUS}

\bibitem{Schael:2006cr}
  S.~Schael {\it et al.} [ALEPH and DELPHI and L3 and OPAL 
  Collaborations and LEP Working Group for Higgs Boson Searches],
  Eur.\ Phys.\ J.\  C {\bf 47} (2006) 547
  [arXiv:hep-ex/0602042].

\bibitem{Chankowski:1997zh}
  P.~H.~Chankowski, J.~R.~Ellis and S.~Pokorski,
  Phys.\ Lett.\  B {\bf 423} (1998) 327
  [arXiv:hep-ph/9712234].

\bibitem{Barbieri:1998uv}
  R.~Barbieri and A.~Strumia,
  Phys.\ Lett.\  B {\bf 433} (1998) 63
  [arXiv:hep-ph/9801353].
  
\bibitem{Kane:1998im}
  G.~L.~Kane and S.~F.~King,
  Phys.\ Lett.\  B {\bf 451} (1999) 113
  [arXiv:hep-ph/9810374].

\bibitem{Giusti:1998gz}
  L.~Giusti, A.~Romanino and A.~Strumia,
  Nucl.\ Phys.\  B {\bf 550} (1999) 3
  [arXiv:hep-ph/9811386].

\bibitem{Maniatis:2009re}
  M.~Maniatis,
  Int.\ J.\ Mod.\ Phys.\  A {\bf 25} (2010) 3505
  [arXiv:0906.0777 [hep-ph]].

\bibitem{Ellwanger:2009dp}
  U.~Ellwanger, C.~Hugonie and A.~M.~Teixeira,
  Phys.\ Rept.\  {\bf 496} (2010) 1\newline
  [arXiv:0910.1785 [hep-ph]].

\bibitem{BasteroGil:2000bw}
  M.~Bastero-Gil, C.~Hugonie, S.~F.~King, D.~P.~Roy and S.~Vempati,
  Phys.\ Lett.\  B {\bf 489} (2000) 359
  [arXiv:hep-ph/0006198].

\bibitem{Dermisek:2007ah}
  R.~Dermisek and J.~F.~Gunion,
  Phys.\ Rev.\  D {\bf 77} (2008) 015013
  [arXiv:0709.2269 [hep-ph]].

\bibitem{Ellwanger:2006rm}
  U.~Ellwanger and C.~Hugonie,
  Mod.\ Phys.\ Lett.\  A {\bf 22} (2007) 1581
  [arXiv:hep-ph/0612133].

\bibitem{Barbieri:2006bg}
  R.~Barbieri, L.~J.~Hall, Y.~Nomura and V.~S.~Rychkov,
  Phys.\ Rev.\  D {\bf 75} (2007) 035007
  [arXiv:hep-ph/0607332].

\bibitem{Dermisek:2005ar}
  R.~Dermisek and J.~F.~Gunion,
  Phys.\ Rev.\ Lett.\  {\bf 95} (2005) 041801
  [arXiv:hep-ph/0502105].

\bibitem{Dermisek:2005gg}
  R.~Dermisek and J.~F.~Gunion,
  Phys.\ Rev.\  D {\bf 73} (2006) 111701
  [arXiv:hep-ph/0510322].

\bibitem{Dermisek:2007yt}
  R.~Dermisek and J.~F.~Gunion,
  Phys.\ Rev.\  D {\bf 76} (2007) 095006
  [arXiv:0705.4387 [hep-ph]].

\bibitem{Dermisek:2009si}
  R.~Dermisek,
  Mod.\ Phys.\ Lett.\  A {\bf 24} (2009) 1631
  [arXiv:0907.0297 [hep-ph]].

\bibitem{Ellis:1986yg}
  J.~R.~Ellis, K.~Enqvist, D.~V.~Nanopoulos and F.~Zwirner,
  Mod.\ Phys.\ Lett.\  A {\bf 1} (1986) 57.

\bibitem{barbieri}
   R. Barbieri and G.~F.~Giudice, Nucl.\ Phys.\ B {\bf 306} (1988) 63.

\bibitem{deCarlos:1993yy}
  B.~de Carlos and J.~A.~Casas,
  Phys.\ Lett.\  B {\bf 309} (1993) 320
  [arXiv:hep-ph/9303291].

\bibitem{Ciafaloni:1996zh}
  P.~Ciafaloni and A.~Strumia,
  Nucl.\ Phys.\  B {\bf 494} (1997) 41
  [arXiv:hep-ph/9611204].

\bibitem{Chankowski:1998xv}
  P.~H.~Chankowski, J.~R.~Ellis, M.~Olechowski and S.~Pokorski,
  Nucl.\ Phys.\  B {\bf 544} (1999) 39
  [arXiv:hep-ph/9808275].

\bibitem{Cassel:2009cx}
  S.~Cassel, D.~M.~Ghilencea and G.~G.~Ross,
  Phys.\ Lett.\  B {\bf 687} (2010) 214
  [arXiv:0911.1134 [hep-ph]].

\bibitem{Cassel:2010px}
  S.~Cassel, D.~M.~Ghilencea and G.~G.~Ross,
  Nucl.\ Phys.\  B {\bf 835} (2010) 110
  [arXiv:1001.3884 [hep-ph]].

\bibitem{Cassel:2011tg}
  S.~Cassel, D.~M.~Ghilencea, S.~Kraml, A.~Lessa and G.~G.~Ross,
  JHEP {\bf 1105} (2011) 120
  [arXiv:1101.4664].

\bibitem{Cassel:2011zd}
  S.~Cassel and D.~M.~Ghilencea,
  ``A review of naturalness and dark matter prediction for the Higgs
  mass in MSSM and beyond,''
  arXiv:1103.4793 [hep-ph].

\bibitem{Anderson:1994dz}
  G.~W.~Anderson and D.~J.~Castano,
  Phys.\ Lett.\  B {\bf 347} (1995) 300
  [arXiv:hep-ph/9409419].

\bibitem{Chan:1997bi}
  K.~L.~Chan, U.~Chattopadhyay and P.~Nath,
  Phys.\ Rev.\  D {\bf 58} (1998) 096004
  [arXiv:hep-ph/9710473].

\bibitem{Strumia:2011dv}
  A.~Strumia,
  JHEP {\bf 1104} (2011) 073
  [arXiv:1101.2195].
  
\bibitem{Akula:2011zq}
  S.~Akula, N.~Chen, D.~Feldman, M.~Liu, Z.~Liu, P.~Nath and G.~Peim,
  Phys.\ Lett.\  B {\bf 699} (2011) 377
  [arXiv:1103.1197].

\bibitem{Conley:2011nn}
  J.~A.~Conley, J.~S.~Gainer, J.~L.~Hewett, M.~P.~Le and T.~G.~Rizzo,
  ``Supersymmetry Without Prejudice at the 7 TeV LHC,''
  arXiv:1103.1697 [hep-ph].

\bibitem{Farina:2011bh}
  M.~Farina, M.~Kadastik, D.~Pappadopulo, J.~Pata, M.~Raidal and A.~Strumia,
  ``Implications of XENON100 results for Dark Matter models and for the LHC,''
  arXiv:1104.3572 [hep-ph].

\bibitem{Sakurai:2011pt}
  K.~Sakurai and K.~Takayama,
  ``Constraint from recent ATLAS results to non-universal sfermion mass
  models and naturalness,''
  arXiv:1106.3794 [hep-ph].

\bibitem{Scopel:2011qt}
  S.~Scopel, S.~Choi, N.~Fornengo and A.~Bottino,
  Phys.\ Rev.\  D {\bf 83} (2011) 095016
  [arXiv:1102.4033].

\bibitem{LopezFogliani:2009np}
  D.~E.~Lopez-Fogliani, L.~Roszkowski, R.~R.~de Austri and T.~A.~Varley,
  Phys.\ Rev.\  D {\bf 80} (2009) 095013
  [arXiv:0906.4911].

\bibitem{Gunion:2011hs}
  J.~F.~Gunion, D.~E.~Lopez-Fogliani, L.~Roszkowski, R.~R.~de Austri and
  T.~A.~Varley, ``Next-to-Minimal Supersymmetric Model Higgs Scenarios
  for Partially Universal GUT Scale Boundary Conditions,''
  arXiv:1105.1195.

\bibitem{Feroz:2011bj}
  F.~Feroz, K.~Cranmer, M.~Hobson, R.~Ruiz de Austri and R.~Trotta,
  JHEP {\bf 1106} (2011) 042
  [arXiv:1101.3296].

\bibitem{Allanach:2011ut}
  B.~C.~Allanach,
  Phys.\ Rev.\  D {\bf 83} (2011) 095019
  [arXiv:1102.3149].

\bibitem{Bechtle:2011dm}
  P.~Bechtle {\it et al.},
  ``What if the LHC does not find supersymmetry in the sqrt(s)=7 TeV
  run?,'' arXiv:1102.4693 [hep-ph].

\bibitem{Allanach:2011wi}
  B.~C.~Allanach, T.~J.~Khoo, C.~G.~Lester and S.~L.~Williams,
  JHEP {\bf 1106} (2011) 035
  [arXiv:1103.0969].

\bibitem{Buchmueller:2011aa}
  O.~Buchmueller {\it et al.},
  Eur.\ Phys.\ J.\  C {\bf 71} (2011) 1634
  [arXiv:1102.4585].

\bibitem{Bechtle:2011it}
  P.~Bechtle {\it et al.},
  ``Present and possible future implications for mSUGRA of the non-discovery of
  SUSY at the LHC,''
  arXiv:1105.5398 [hep-ph].

\bibitem{Ross:1992tz}
  G.~G.~Ross and R.~G.~Roberts,
  Nucl.\ Phys.\  B {\bf 377} (1992) 571.

\bibitem{Ellwanger:2006rn}
  U.~Ellwanger and C.~Hugonie,
  Comput.\ Phys.\ Commun.\  {\bf 177} (2007) 399
  [arXiv:hep-ph/0612134].

\bibitem{Ellwanger:2004xm}
  U.~Ellwanger, J.~F.~Gunion and C.~Hugonie,
  JHEP {\bf 0502} (2005) 066
  [arXiv:hep-ph/0406215].

\bibitem{Ellwanger:2005dv}
  U.~Ellwanger and C.~Hugonie,
  Comput.\ Phys.\ Commun.\  {\bf 175} (2006) 290
  [arXiv:hep-ph/0508022].

\bibitem{Ellwanger:2010es}
  U.~Ellwanger, A.~Florent and D.~Zerwas,
  JHEP {\bf 1101} (2011) 103
  [arXiv:1011.0931 [hep-ph]].

\bibitem{Abbiendi:2004ww}
  G.~Abbiendi {\it et al.}  [OPAL Collaboration],
  Eur.\ Phys.\ J.\  C {\bf 37} (2004) 49
  [arXiv:hep-ex/0406057].
  
\bibitem{Abdallah:2004wy}
  J.~Abdallah {\it et al.}  [DELPHI Collaboration],
  Eur.\ Phys.\ J.\  C {\bf 38} (2004) 1
  [arXiv:hep-ex/0410017].

\bibitem{Schael:2010aw}
  S.~Schael {\it et al.}  [ALEPH Collaboration],
  JHEP {\bf 1005} (2010) 049
  [arXiv:1003.0705].

\bibitem{Domingo:2009tb}
  F.~Domingo, U.~Ellwanger and M.~A.~Sanchis-Lozano,
  Phys.\ Rev.\ Lett.\  {\bf 103} (2009) 111802
  [arXiv:0907.0348 [hep-ph]].

\bibitem{Domingo:2011rn}
  F.~Domingo and U.~Ellwanger,
  JHEP {\bf 1106} (2011) 067
  [arXiv:1105.1722].

\bibitem{Ellwanger:2008py}
  U.~Ellwanger, C.~C.~Jean-Louis and A.~M.~Teixeira,
  JHEP {\bf 0805} (2008) 044
  [arXiv:0803.2962].

\end{thebibliography}
\end{document}